\documentclass[letter, longauth]{aa} 

\usepackage{graphicx}
\usepackage{txfonts}
\usepackage{xcolor}

\newcommand{\HeI}{\mbox{He\,{\sc i}}}

\newcommand{\CII}{\mbox{C\,{\sc ii}}}

\newcommand{\OII}{\mbox{O\,{\sc ii}}}
\newcommand{\SiII}{\mbox{Si\,{\sc ii}}}
\newcommand{\SiIII}{\mbox{Si\,{\sc iii}}}

\newcommand{\SII}{\mbox{S\,{\sc ii}}}
\newcommand{\FeII}{\mbox{Fe\,{\sc ii}}}
\newcommand{\MgII}{\mbox{Mg\,{\sc ii}}}

\newcommand{\vsini}{\mbox{$v\sin i$}}
\newcommand{\Teff}{\mbox{$T_{\rm eff}$}}
\newcommand{\logg}{\mbox{$\log g$}}
\newcommand{\logd}{\mbox{$\log d$}}
\newcommand{\MSol}{\mbox{M$_\odot$}}
\newcommand{\RSol}{\mbox{R$_\odot$}}
\newcommand{\LSol}{\mbox{L$_\odot$}}
\newcommand{\kms}{\mbox{km\,s$^{-1}$}}

\begin{document}

   \title{A detailed non-LTE analysis of LB-1: 
   Revised parameters and surface abundances}


   \author{S. Sim\'on-D\'iaz\inst{1,2},
          J. Ma\'iz Apell\'aniz\inst{3},
          D.~J. Lennon\inst{1,2},
          J.~I. Gonz\'alez Hern\'andez\inst{1,2},
          C. Allende Prieto\inst{1,2},
          N. Castro\inst{4}, \newline
          A. de Burgos\inst{5},
          P. L. Dufton\inst{6},
          A. Herrero\inst{1,2},
          B. Toledo-Padr\'on\inst{1,2}
          S. J. Smartt\inst{6}
          }

   \institute{Instituto de Astrof\'isica de Canarias,E-38\,200 La Laguna, Tenerife, Spain
         \and
             Universidad de La Laguna, Universidad de La Laguna, E-38\,205 La Laguna, Tenerife, Spain
         \and
             Centro de Astrobiolog\'ia, ESAC campus, Villanueva de la Ca\~nada, E-28\,692, Spain
         \and
             Leibniz-Institut f\"ur Astrophysik Potsdam (AIP), An der Sternwarte 16, 14\,482 Potsdam, Germany
          \and
             Nordic Optical Telescope, Rambla Jos\'e Ana Fern\'andez P\'erez 7, E-38\,711 Bre\~na Baja, Spain
          \and
             Astrophysics Research Centre, School of Mathematics \& Physics,  Queen’s University, Belfast, BT7 1NN, UK 
             }
         
   \titlerunning{Revised parameters and surface abundances of LB-1}
   \authorrunning{Sim\'on-D\'iaz et al.}
   
   \date{Received xxx; accepted xxx}

 
  \abstract
   {LB-1 has recently been proposed to be a binary system at 4~kpc consisting of a B-type star of 8~\MSol\ and a massive stellar black hole (BH) of 70 \MSol. This finding challenges our current theories of massive star evolution and formation of BHs at solar metallicity.}
   {Our objective is to derive the effective temperature, surface gravity and chemical composition of the B-type component in order to determine its nature and evolutionary status and, indirectly, to constrain the mass of the BH.}
   {We use the non-LTE stellar atmosphere code {\sc FASTWIND} to analyse new and archival high resolution data.} 
   {We determine (\Teff, \logg) values of (14\,000$\pm500$\,K, 3.50$\pm0.15$ dex) that, combined with the {\it Gaia} parallax, implies a $spectroscopic$ mass, from \logg, of $3.2^{+2.1}_{-1.9}$~\MSol\ and an $evolutionary$ mass, assuming single star evolution, of $5.2^{+0.3}_{-0.6}$ 
   \MSol. We determine an upper limit of 8~\kms\ for the projected rotational velocity and derive the surface abundances, finding the star to have a silicon abundance below solar, to be significantly enhanced in nitrogen and iron, and depleted in carbon and magnesium. Complementary evidence derived from a photometric extinction analysis and {\it Gaia} yields similar results for \Teff\ and \logg\ and a consistent distance around 2~kpc.  
   }
   {We propose that the B-type star is a slightly evolved main sequence star of 3\,--\,5~\MSol\ with surface abundances reminiscent of diffusion in late B/A chemically peculiar stars with low rotational velocities. There is also evidence for CN-processed material in its atmosphere. These conclusions rely critically on the distance inferred from the {\it Gaia} parallax. The goodness of fit of the {\it Gaia} astrometry also favours a high-inclination orbit. If the orbit is edge-on and the B-type star has a mass of 3\,--\,5~\MSol, the mass of the dark companion would be 4\,--\,5~\MSol, which would be easier to explain with our current stellar evolutionary models.}

   \keywords{techniques: spectroscopic, binaries: spectroscopic, stars: black holes, stars: early-type, stars: fundamental parameters, stars: abundances}

   \maketitle
%

\section{Introduction}


$\,\!$\indent The LAMOST collaboration has recently announced the potential discovery of a black-hole (BH) in the system LB-1 (LS~V~+22~25, ALS~8775) of about 70 \MSol\ \citep{Liuetal19}. The object is part of a wide binary system in which the visible component is an extremely narrow lined B-type star. The mass determination of the dark companion is based on three key empirical measurements. First, the radial velocity curve of the B-type star, which allowed identification of the period  (P\,=\,78.9\,$\pm$\,0.3~d) and the mass function ($PK_{\rm B}/2\pi G$\,=\,1.20\,$\pm$\,0.05\,\MSol, with $K_{\rm B}$\,=\,52.8\,$\pm$\,0.7~\kms) of the binary system. Second, the mass of the B-type star, which was proposed to be 8.2$^{+0.9}_{-1.2}$\,\MSol. Finally, the authors measure changes in radial velocity associated with the broad and strong H${\alpha}$ emission line which they suggest are associated with a Keplerian disc orbiting the black hole. If correct this discovery would have important implications for stellar evolution \citep{groh,belczynski,shen}.

Since then, however, it has been shown that the radial velocity measurements of the H$\alpha$ emission line are spurious \citep{elbadry,abdulmasih}, resulting from the influence of the underlying broad stellar absorption typical of late main-sequence B-type stars. This was not taken into account by \citet{Liuetal19}. \citet{eldridge} further propose, based on BPASS binary evolution models, that a more likely scenario is a BH mass around 4\,--\,7~\MSol, and that the distance implied by the $Gaia$ parallax is correct. This distance is almost a factor of two smaller than that inferred by \citeauthor{Liuetal19} when assuming that the mass of the B-star was 8.2~\MSol. 

In this letter we focus on a detailed non-LTE quantitative spectroscopic analysis of a set of high quality spectra to determine the stellar parameters, including the so-called evolutionary and spectroscopic masses, and chemical composition of the B-type star. Our ultimate aim is to check for consistency with the currently evolving picture of LB-1 as a typical, though extremely interesting, B+BH (B-type star plus black hole) system at a distance of approximately 2\,kpc.  In section 2 we discuss the observational data.
In section 3 we present our analysis and results, and in section 4 we discuss some implications.

\section{Observations}

$\,\!$\indent On November 4 and 5, 2019, we obtained two 1800\,s exposures with the HARPS-N@TNG3.5~m \'echelle spectrograph. Spectral resolution ($R$) is 115\,000 with wavelength coverage from 383 to 690~nm.  The signal-to-noise ratio (S/N) per pixel in the blue spectral ranges from 35 to 70 in the co-added and re-binned (0.05~\AA/pixel) spectrum. 

On November 30 2019, we also obtained a set of five 15-minute, $R$\,=\,25\,000 spectra with the HORuS@GTC10.4~m echelle spectrograph, which provides nearly complete spectral coverage between 380 and 690 nm. The combined HORuS spectrum\footnote{Individual spectra were reduced with the HORuS chain: github.com/callendeprieto/chain.} has a median S/N per pixel (1/3 of a resolution element) of 150 (210 at 500 nm). 

We also downloaded the raw HIRES ($R$$\sim$50\,000) data presented by \citet{Liuetal19} from the Keck archive and used standard {\sc IRAF} procedures to re-reduce the 600s single exposure from 24 December 2017 (the highest S/N spectrum from the HIRES dataset).  
During the reduction process, no flat-fielding was applied since this does not improve the S/N of the extracted spectrum and causes further problems in normalising the orders which contain broad Balmer lines.   

Last, we downloaded all 12 publicly available LAMOST spectra from DR5v3. The standard flux calibrated pipeline products were normalized for comparison with the other high resolution, non flux calibrated spectra (see below and Appendix~\ref{app_balmer}).

In comparing model spectra and observed line profiles all data must be appropriately normalized. However, the blaze correction and normalization of those individual orders in an \'echelle spectrum that includes broad and extended lines (as is the case of the Balmer lines in late B-type or A-type stars) is difficult. Even a small misplacement of the continuum can significantly distort the shape of the far line-wings, that are the most reliable indicators of surface gravity. This reliance on the line wings is even more important for LB-1 as there is clearly emission effecting the intensity of the Balmer line cores. We refer the reader to Appendix\,\ref{app_balmer} for a more technical discussion of this aspect of our analysis.

\section{Stellar parameters and surface abundances of the B-type star in LB-1 }

   \begin{figure}[!t]
   \centering
   \includegraphics[width=0.47\textwidth, angle=0]{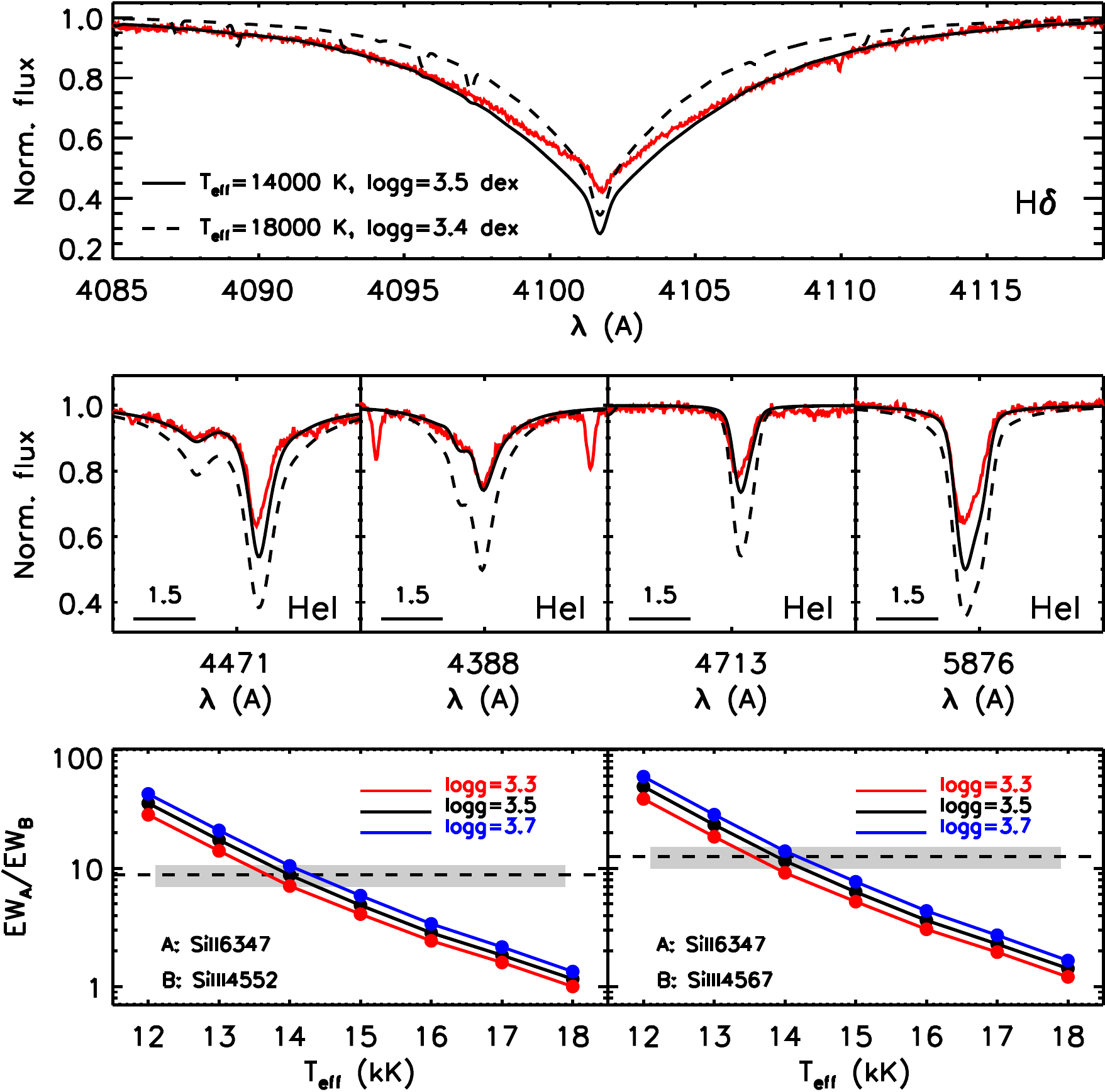}
      \caption{Summary of the quantitative spectroscopic analysis strategy we follow for the determination of \Teff\ and \logg. Top and middle panels show line profiles of H${\delta}$ and some \HeI\ lines in the HIRES spectrum, respectively. Solid and dashed lines in these panels depict the synthetic spectra associated with two {\sc FASTWIND} models with a different (\Teff, \logg) pair: our best solution (14\,000~K, 3.5~dex, solid) and the one proposed by \citet{Liuetal19} (18\,000~K, 3.4~dex, dashed). Note that H$\delta$ and some \HeI\ lines show infilling from the disk emission. Bottom panels show the predicted variation of the \SiII\,$\lambda$6347/\SiIII\,$\lambda$4552 (left) and \SiII\,$\lambda$6347/\SiIII\,$\lambda$4567 (right) ratios with \Teff\ for three values of \logg\ (Si abundance fixed to solar). The horizontal grey band indicates the empirical measurement and its associated uncertainty.}
              \label{FigDel}%
    \end{figure}

\subsection{Spectroscopic parameters}\label{analysis}

$\,\!$\indent As a first step in our analysis we used a grid of standards for spectral classification of OB stars to determine the spectral type and luminosity class of LB-1 (see Appendix\,\ref{app_classify}). We obtain B6~IVe, a later spectral type that implies a substantially cooler effective temperature than that obtained by \citet{Liuetal19}. We also checked in the three high resolution spectra for signatures of a hidden fast rotating B-type star which could be associated with the strong H$_{\alpha}$ emission thorugh a Be star phenomenon, but we did not find any.

To perform the quantitative spectroscopic analysis, including the determination of the projected rotational velocity (\vsini), the spectroscopic parameters (\Teff, \logg) and the abundance analysis, we used the {\sc iacob-broad} tool \citep{simondiaz2014} and a grid of synthetic spectra obtained with the non-LTE, line-blanketed, spherical stellar atmosphere code {\sc FASTWIND} \citep{puls2005}. 

The {\sc iacob-broad} analysis of the \SiII\,$\lambda$6371\,\AA\ line leads to an upper limit in \vsini\ of 8~\kms, a result that is also supported by the fact that the two components of the \MgII\,$\lambda\lambda$4481.126,4481.325\,\AA\ doublet can be resolved in the HARPS-N spectrum. The analysis of other lines gives a similar result. In all cases we find that the profiles cannot be properly fitted with a pure rotational profile, indicating that there is some additional broadening affecting the line, and hampering a proper determination of \vsini\ below the proposed value.

   \begin{figure*}[t!]
   \centering
   \includegraphics[width=0.8\textwidth]{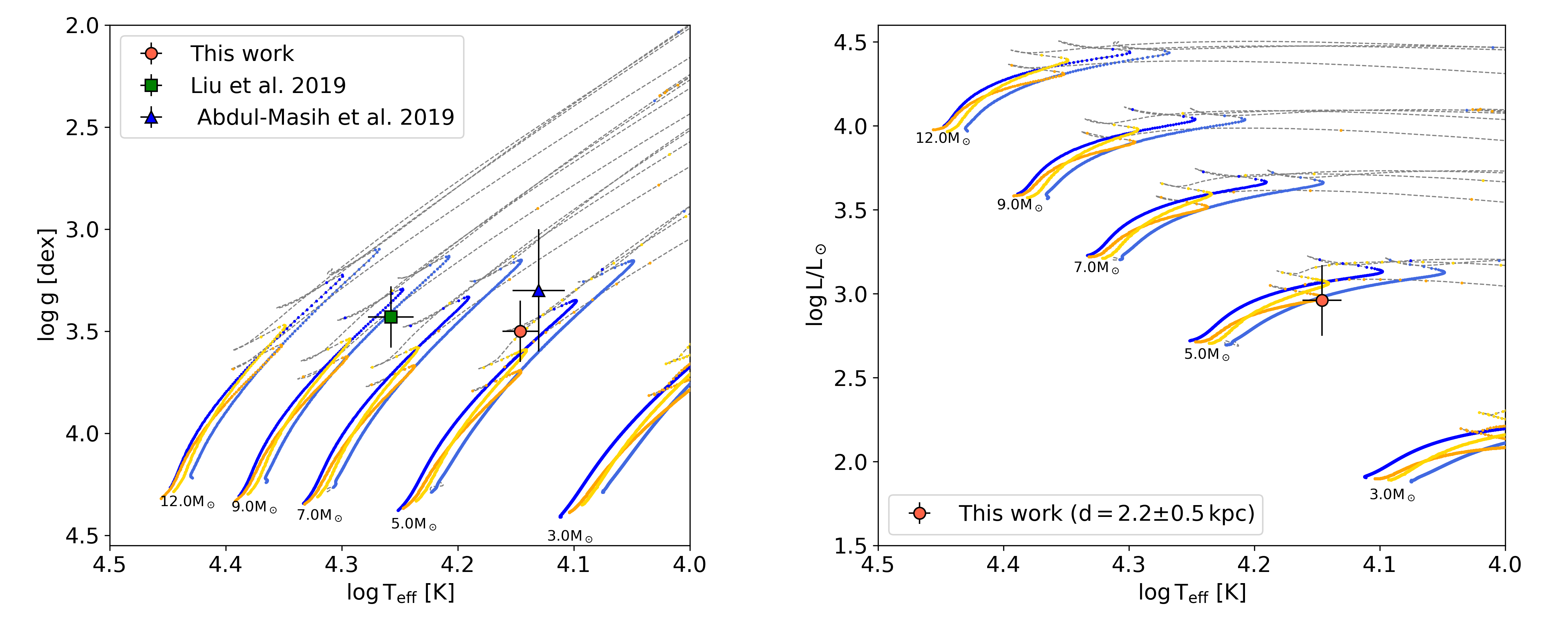}
   \caption{Location of the B star component of the LB-1 system in the Kiel (left) and HR (right) diagrams. The various evolutionary tracks correspond to models with and without rotation from \citet[][orange]{ekstrom2012} and \citet[][blue]{brott}. In the Kiel diagram we also indicate the location of the star accordingly to the parameters derived by \citet{Liuetal19} and \citet{abdulmasih}. In the HR diagram we show the position of the star assuming the Gaia distance computed in Appendix~\ref{app_gaia}.}
              \label{kiel+hrd}%
    \end{figure*}

\begin{table*}[!t]
\caption{Summary of spectroscopic and fundamental parameters resulting from the quantitative spectroscopic analysis of the late-B star in the LB-1 system. Stellar radius, luminosity and spectroscopic mass have been computed assuming a distance to the star $d$\,=\,2.2$\pm$0.5~kpc, and an extinction $A_V$\,=\,1.503$\pm$0.027~mag. Evolutionary mass obtained with {\sc BONNSAI}
\citep{schneider2014} using \Teff, \logg, and log~$L$ as input parameters. Abundances have been obtained assuming fixed values of \Teff\ (14\,000 K), \logg\ (3.5~dex), helium abundance (N$_{\rm He}$/N$_{\rm H}$\,=\,0.1), and microturbulence (1~\kms). Last column refers to the abundances derived in B-type stars in the Orion~OB1 association \citet{NievaSimon11}, which can be considered as the present-day abundances of the solar neighborhood up to a few kpc.}
\label{table:1}
\centering
\begin{tabular}{rlrrlrrlrrr}
\hline \hline
            \noalign{\smallskip}
            \multicolumn{3}{c}{Spectroscopic parameters} & \multicolumn{3}{c}{Fundamental parameters} & \multicolumn{3}{c}{Surface abundances (LB-1)} & \multicolumn{2}{c}{Ref. abund.} \\
            \noalign{\smallskip}
            \hline
            \noalign{\smallskip}
            \Teff\,=          & 14\,000 $\pm$500    & [K]    & $R$\,=                  & 5.3 $\pm$ 1.2    & [\RSol]   & $\epsilon$(Si)\,= & 7.20 $\pm$ 0.15 & [dex]  &  7.50$\pm$0.06 & [dex] \\
            \logg\,=        & 3.5 $\pm$ 0.15     & [dex]  & log($L$/\LSol)\,= & 2.98 $\pm$ 0.21  & [dex]               & $\epsilon$(Mg)\,= & 6.87 $\pm$ 0.15 &  [dex] & 7.57$\pm$0.06 & [dex] \\
            $v$~sin$i$\,=    & $<$ 10       & [km/s] & $M_{\rm sp}$\,=         & 3.2$^{+2.1}_{-1.9}$    & [\MSol]      & $\epsilon$(C)\,=  & 7.50 $\pm$ 0.2 &  [dex]  & 8.35$\pm$0.03  & [dex]\\
            Y$_{\rm He}$\,=  & 0.1          &        & $M_{\rm ev}$\,=         & 5.2$^{+0.3}_{-0.6}$        & [\MSol]    & $\epsilon$(N)\,=  &  8.20$\pm$0.15 &  [dex]  & 7.82$\pm$0.07  & [dex]\\
            \noalign{\smallskip}
            \hline
\end{tabular}
\end{table*}

Due to the very low \vsini\ of LB-1, and the high S/N of the data, we can securely measure extremely weak lines such as the \SiIII\,$\lambda\lambda$4552,4567\,\AA\ lines, that have equivalent widths of roughly 11 and 8~m\AA\ respectively (correcting the former for a blend with a \SII\ line). Hence, we are able to use the \SiII/\SiIII\ ionization equilibrium (together with the fitting of the wings of some of the Balmer lines, see below) to estimate the effective temperature of the star (see Fig.~\ref{FigDel}). Our estimate for this parameter is 14~000\,$\pm$\,500~K, which is in fairly good agreement with the value obtained by \citet{abdulmasih}, but 4\,000~K less than that proposed by \citet{Liuetal19}.

In addition, several \HeI\ lines are used as a consistency check of the temperature as well as to have information about the helium abundance, having in mind that some of these lines (like \HeI\,$\lambda\lambda$4471 and 5875~\AA, see middle panels in Fig.~\ref{FigDel}) may be contaminated by emission from the disk, whose nature -- circumbinary or circumstellar -- is still unknown \citep{elbadry,abdulmasih}. The good fit to the other \HeI\ lines for the derived \Teff\ indicates that the helium abundance of the star is N(He)/N(H)\,=\,0.1 ($i.e.$ normal).

Gravity is constrained at the same time as \Teff\ using the wings of the Balmer lines, specially H$\delta$ and H$\gamma$, since H$\beta$ and H$\alpha$ are significantly contaminated by the strong disc emission. The derived value of \logg\ is 3.50$\pm$0.15 dex and we show our final fit of the H$_{\delta}$ profile in Fig.~\ref{FigDel}. We also compare a {\sc FASTWIND} model for the \Teff\ and \logg\ determined by \citet{Liuetal19}. As can be seen, the combination of effective temperature and gravity estimated by these authors results in a poor fit to the wings of the Balmer lines.
The reason is likely due to some form of processing (e.g. flat-fielding and illumination corrections combined with the strong blaze function of the orders) of the Keck spectrum which significantly affects the normalization of the Balmer lines if not done carefully (see Appendix\,\ref{app_balmer}), plus their overestimate of \Teff\, which is too high to fit the \SiIII/\SiII\ ionization equilibrium.

\subsection{Fundamental parameters}

$\,\!$\indent By locating the star in the Kiel diagram (Fig.~\ref{kiel+hrd}, left) and comparing with different evolutionary tracks for massive single stars \citep{ekstrom2012,brott} we derive an evolutionary mass\footnote{This mass estimate assumes that the star is on (or close to) the main sequence and its evolution has not been affected by its companion, which might not be necessarily the case \citep[see][]{eldridge}.} ($M_{\rm ev}$) of $\sim$5~\MSol. This should be compared with the value 8.2~\MSol\ from \citeauthor{Liuetal19} and the lower estimate of 4.2$^{+0.8}_{-0.7}$ \MSol from \citet{abdulmasih}. The main reason of this reduction by 40-50\% from the evolutionary mass estimation by \citet{Liuetal19} is the lower \Teff\ obtained in this work, and in \citet{abdulmasih}.

   \begin{figure*}[!t]
   \includegraphics[width=0.19\textwidth, angle=90]{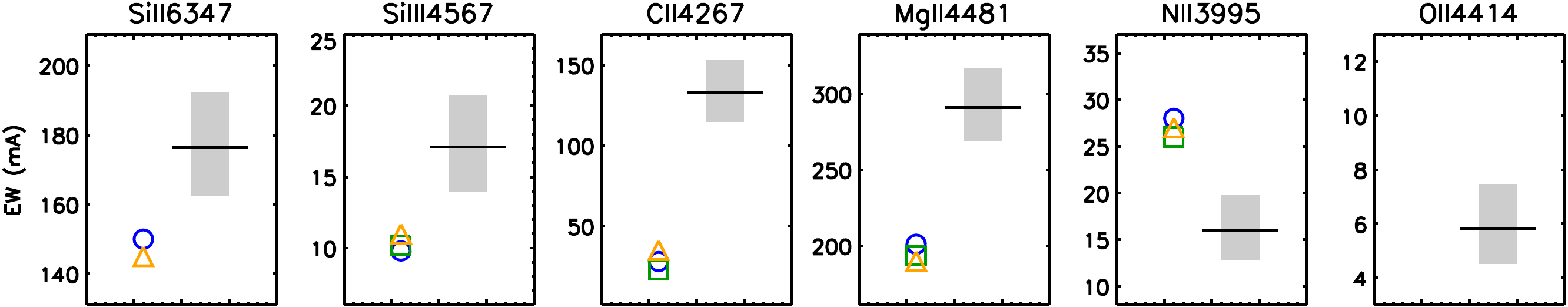}
   \caption{Open symbols: $EW$ measurements for the indicated diagnostic lines using the various high resolution spectra available (circles: HIRES; triangles: HARPS-N; squares: HORuS). Grey rectangles: Predicted range of $EW$s from {\sc FASTWIND} models with the same \Teff\ and \logg\ and a range in abundance $\pm$0.15~dex around the reference value (solid horizontal line, see Table 1). Right panel, including the predicted EW of the \OII$\lambda$4414~\AA\ has been included for reference, to indicate that the non detection of \OII\ lines does not necessarily imply a low oxygen abundance.}
             \label{EWs}%
    \end{figure*}

An alternative estimate of the mass of the B-type star, its $spectroscopic$ mass, can be derived from \logg\ and the stellar radius ($R$), the later being determined from the observed spectral energy distribution ({\rm  SED}) and distance. 
Assuming the distance derived from the $Gaia$ parallax (2.20$^{+0.49}_{-0.35}$~kpc, see Appendix D) and using the {\sc IDL} tool {\sc CHORIZOS} to estimate the extinction ($A_{\rm V}$\,=\,$1.503\pm 0.027$ mag, see Appendix\,\ref{app_chorizos}) we obtain an absolute visual magnitude $M_V$\,=\,-1.6$\pm$0.5 mag. Note that the dominant uncertainty is the one coming from the distance. By fitting the synthetic {\sc SED} of the best fitting {\sc FASTWIND} model to this value of $M_{\rm v}$, we derive a radius of 5.3$\pm$1.2~\RSol\ and, hence, using our derived \Teff\ and \logg, a luminosity $\log L$/\LSol\,=2.98\,$\pm$\,0.21 and a spectroscopic mass $M_{\rm sp}$\,=\,$3.2^{+2.1}_{-1.9}$~\MSol.

We can then use the derived luminosity to locate the star in the HR diagram (Fig.~\ref{kiel+hrd}, right). In this case, the derived evolutionary mass is $\sim$\,5~\MSol, similar to that obtained from the Kiel diagram.
Clearly there is a mismatch between evolutionary/Kiel masses and the spectroscopic mass, though we can rule out much lower masses, such as might be relevant to a post-AGB scenario as that would require a much lower surface gravity, inconsistent with the observations. However we have implicitly assumed a normal single star evolutionary scenario, which may not be valid for this object. An analysis with CHORIZOS where \Teff\ is derived from the photometry yields similar results (Appendix C).

As discussed in more detail in Appendix D, this range of masses for the B-type star, together with the apparent lack of motion in $Gaia$, are all consistent with a hypothesis of a high inclination system, implying a BH mass of only 3.8-4.9 \MSol.  In that case the B-type star either has its rotation axis almost perpendicular to its orbital axis, or its rotational velocity is intrinsically very small. 

\subsection{Surface abundances}

$\,\!$\indent Due to the high S/N of our the data and low \vsini\ of the star we can reliably measure equivalent widths ($EW$s) for lines of 10 m\AA\ or less. 
Figure~\ref{EWs} summarizes the $EW$ measurements for a representative set of lines of the various ions indicated above for which we were able to determine abundances (i.e., those elements included in our {\sc FASTWIND} computations).
The figure presents a comparison of the empirical $EW$s with those predicted by our {\sc FASTWIND} grid of models for fixed values of effective temperature (14\,000~K), surface gravity (3.5~dex), helium abundance (N$_{\rm He}$/N$_{\rm H}$\,=\,0.1), and microturbulence (1~\kms) and abundances covering a range $\pm$0.15~dex around the reference abundances, corresponding to B-type stars in the Orion~OB1 association (see Table~\ref{table:1}). The narrowness of the metal lines constrains the microturbulence to less than 3~\kms\ but we note that the abundance estimates are sensitive to the adopted value; for example increasing the microturbulence from 1 to 3~\kms, decreases the N abundance estimates by 0.1\,--\,0.2~dex. 

The outcome of this abundance analysis is summarized in Table~\ref{table:1}. In brief, we find that (1) the star has a silicon abundance that is 0.3 dex below the reference one; (2) carbon and magnesium surface abundances are depleted by more than a factor 7 and 5, respectively; and (3) nitrogen is enhanced by a factor $\sim$2.5. In addition, we note that, despite the high S/N ratio of the HIRES and HORuS spectra, we have not been able to detect the \OII$\lambda$4414~\AA\ line, one of the strongest \OII\ lines in the optical spectrum; however, this does not necessarily imply a low oxygen abundance provided the small EW predicted by {\sc FASTWIND} (see Fig.~\ref{EWs}). Last, the {\sc FASTWIND} synthetic spectra do not include iron lines. However, comparison of the observed \FeII\ spectrum with those predicted by TLUSTY models in the Bstar grid  \citep{LanzHube07} implies that iron is enhanced in this star by $\sim$~0.15~dex\footnote{We have checked that the change in Si abundance or overall metallicity have no significant impact on Teff and log g, that were initially determined with solar abundances.}.

\section{Concluding remarks}

$\,\!$\indent There is no doubt that the spectrum of the B-type star is peculiar in a number of ways. Its projected rotational velocity is very small, $\vsini$< 8~\kms, and if we accept that the high inclination angle implied by the disk applies to the B-star then the stellar rotational velocity is also very small. 
If the BH is the result of an interacting binary evolution scenario \citep{eldridge} then such a scenario must explain the small rotational velocity. 
The surface composition is also peculiar. The overabundance of N and depletion of C implies contamination with CN processed material, yet the depletion of Mg and and enhancement of Fe are reminiscent of diffusion abundance patterns that are common in late B-type stars with low rotational velocities \citep{hempel}.
The system is also rather isolated and while its proper motion properties are normal, the systemic radial velocity (+28.9\,\kms) implies an LSR velocity of approximately +16.4\,\kms \citep[using the solar peculiar motion from][]{schonrich}. 
The star is almost directly anti-centre so this is mildly peculiar, for example direct measurements of massive star forming regions in this part of the Milky Way have LSR velocities of less than $\pm 10$ \kms\ \citep{reid}. 
This peculiar velocity may be the signature of a kick from the formation of the BH.  
Finally it is strange that there are no apparent signs of the motion of the B star around the dark companion in the $Gaia$ data, and we propose a hypothesis that requires a near edge-on orbit, in which case the dark companion would have a mass as low as 4-5~\MSol, an order of magnitude lower than the \citet{Liuetal19} value (Appendix D).

\begin{acknowledgements}
      We thank J. Hern\'andez for his help with {\it Gaia}~DR2 data.
      S-SD, DJL, AdB, AHD and JMA acknowledges support from the Spanish Government Ministerio de Ciencia, Innovaci\'on y Universidades through grants PGC2018-095\,049-B-C22 and PGC-2018-091\,3741-B-C22. JIGH acknowledges financial support from the Spanish Ministry of Science, Innovation and Universities (MICIU) under the 2013 Ram\'on y Cajal program RYC-2013-14875.  JIGH and CAP acknowledge financial support from the Spanish Ministry project MICIU AYA2017-86389-P. SJS acknowledges funding from STFC Grant ST/P000312/1. Based on observations made with the Telescopio Nationale Galileo (TNG) and the Gran Telescopio Canarias (GTC), installed at the Spanish Observatorio del Roque de los Muchachos of the Instituto de Astrof\'sica de Canarias, in the island of La Palma. This research has made use of the Keck Observatory Archive (KOA), which is operated by the W. M. Keck Observatory and the NASA Exoplanet Science Institute (NExScI), under contract with the National Aeronautics and Space Administration. Guoshoujing Telescope (the Large Sky Area Multi-Object Fiber Spectroscopic Telescope LAMOST) is a National Major Scientific Project built by the Chinese Academy of Sciences. Funding for the project has been provided by the National Development and Reform Commission. LAMOST is operated and managed by the National Astronomical Observatories, Chinese Academy of Sciences.  

\end{acknowledgements}

%
%




\begin{appendix}

\section{Balmer line profiles}
\label{app_balmer}

\begin{figure}
\centerline{\includegraphics*[width=\linewidth]{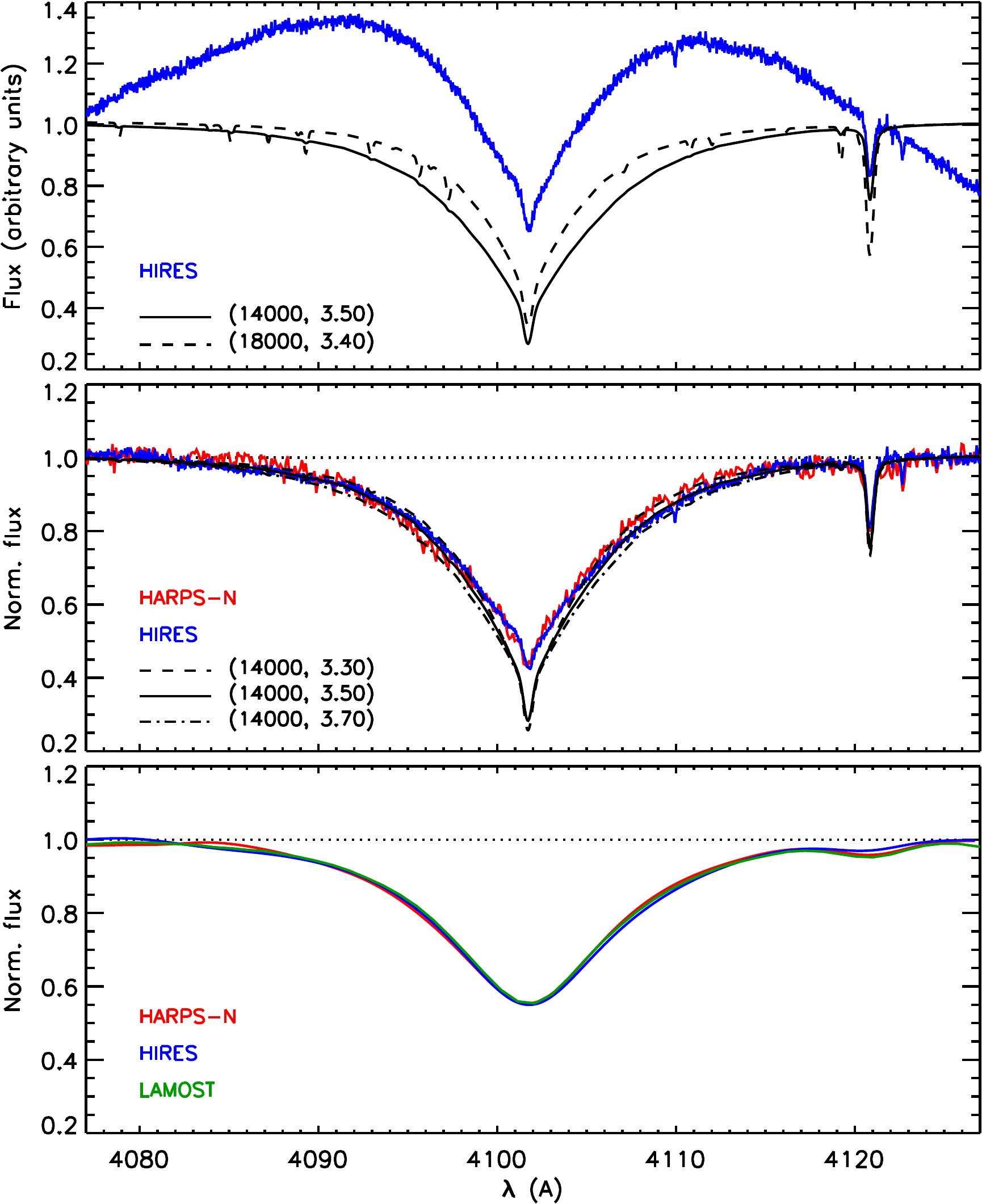}}
\caption{Top panel: The blue line is the scaled extracted echelle order from Keck/HIRES, the black continuous and dashed lines are the {\sc FASTWIND} profiles for the indicated parameters. Middle panel: Blue and red lines are the line profiles normalized using the procedure discussed in the text of Appendix A, while the black lines illustrate the effect of a change of $\pm$0.2~dex in \logg\ at constant \Teff. Bottom panel: Here we compare the smoothed, normalized profiles for TNG/HARPS-N, Keck/HIRES and LAMOST, illustrating the good consistency obtained for all three datasets.}
\label{hdelta}
\end{figure}

$\,\!$\indent A major difference in the present analysis compared to \citet{Liuetal19} is the determination of parameters that imply a significantly lower mass, provided the B-star is a normal main sequence object. 
In the course of this work we noted an apparent difference between the wings of the Balmer lines in Fig.1b of \citet{Liuetal19}, obtained from Keck HIRES data, and those presented here. In particular we note that their H$\delta$ line profile appears to much narrower than in our data. This is most easily seen by comparing the red wing of H$\delta$ around the region of the He\,{\sc i} line at 4121\AA\ (middle panel of Fig.~\ref{hdelta}).
While the good agreement between our various observations shown in Fig.~\ref{hdelta} lends support to our data reduction processes it is worth recalling the difficulties entailed in normalizing the Hydrogen Balmer lines, in particular as observed with echelle spectrographs.
In the relevant temperature and gravity range the pressure sensitive wings of these lines are extremely broad, and cover all of an echelle order in Keck/HIRES\footnote{Or even extend into two consecutive orders, as is the case of the HARPS-N spectrum, for which the boundary between the two orders is located $\sim$4085~\AA, making it even more difficult the blaze correction and normalization process.}, as shown in the upper panel of Fig.~\ref{hdelta}.
This order contains H$\delta$ and extends from approximately 4060 to 4134 \AA, and at no point in this interval does the theoretical profile reach the continuum for any of the models discussed in this paper. 
At best they reach 99\% of the continuum and at worst 97.5\%. 
Therefore if normalization of the profile is performed within this spectral range then care must be taken to treat the theoretical profiles in a consistent manner. 
In order to derive normalized Balmer line profiles from the Keck/HIRES data we first approximate the blaze function by interpolating between neighbouring orders (in pixel space). The result is an approximately normalized order and any residual slope is removed by fitting a straight line through line-free regions, being careful not to define continuum points within $\pm$~20~\AA\ of line center. Theoretical Balmer line profiles are normalized in the same way (cf Fig.~\ref{hdelta}, middle panel). 

\section{Spectral classification}
\label{app_classify}

$\,\!$\indent We review the spectral classification of the B star by first degrading the HARPS-N spectrum to the typical classification resolution of $R\sim2500$, which is very important for such a sharp lined star. We 
compare a normalized spectrum with a grid of spectral classification standards, in this case a blue-violet 
grid 
developed with data from the Galactic O-Star Spectroscopic Survey (GOSSS, \citealt{Maizetal11}) that includes standards in the  spectral type range O2-A0 and in the luminosity class range Vz-Ia+. A comparison with the low-rotation B6~V standard $\beta$~Sex reveals nearly identical \HeI\,$\lambda$4471 and \MgII\,$\lambda$4481 profiles and similar \HeI\ and \CII\ lines in general, establishing the spectral subtype at B6. On the other hand, the H$\gamma$ and H$\delta$ profiles are narrower than those of $\beta$~Sex and similar to those of the B6~IV standard, leading to a luminosity class of IV. LB-1 shows an obvious emission component in the H$\beta$ profile and the metal lines are very narrow (even slightly narrower than those of $\beta$~Sex). There are no other strong differences with $\beta$-Sex (other than the DIBs associated with the higher extinction toward LB-1). Therefore, the spectral type is B6~IVe, significantly later than the B3~V classification proposed by \citet{Liuetal19}. We also note that their luminosity class of V is incompatible with their preferred $\logg = 3.3$, as B3~V stars have values around 4.0.

\section{CHORIZOS analysis}                          
\label{app_chorizos}

$\,\!$\indent We have used the latest version of the CHORIZOS \citep{Maiz04c} photometry-fitting code to derive some of the parameters
of LB-1. We have done two sets of runs and we start describing the first one.

{\it Photometric data.} For the first set of runs we collected the magnitudes from {\it Gaia}~DR2 
$G+G_{\rm BP}+G_{\rm RP}$ (\citealt{Evanetal18} with the corrections and calibration from \citealt{MaizWeil18}),
APASS $B+V$ (\citealt{Hendetal15} with the calibration from 
\citealt{Maiz06a}), 
2MASS $J+H+K_{\rm s}$
(\citealt{Skruetal06} with the calibration from \citealt{MaizPant18}), and WISE $W1+W2+W3+W4$ \citep{Cutretal13}.
As the object has a significant IR excess, we eliminate from the runs the five photometric points with the 
longest effective wavelengths ($K_{\rm s}$ and the four WISE bands) but include them in the analysis to evaluate 
the excess.

{\it Models.} We use a \Teff-luminosity class grid with solar metallicity 
from \citet{Maiz13a}. The 
temperature of interest lies at the transition region between the TLUSTY \citep{LanzHube07} and Munari 
\citep{Munaetal05} parts of the grid, so data from both sources are used. In particular, note that the NIR colors
from TLUSTY models for late/mid-B stars are incorrect by up to several hundredths of a magnitude, so in that region
of the spectrum the grid uses the Munari models for all hot stars.

{\it Parameters.} For the first set of runs we fix \Teff\ to 14~kK from the results in this paper and we do
four individual runs for different values of the luminosity class: 5.5 (ZAMS), 5.0 (typical class V), 4.5 
(luminous class V), and 4.0 (typical class IV). The corresponding values of \logg\ (in cgs) are 4.27, 4.04, 
3.59, and 3.38; of the initial stellar mass (in \MSol) are 3.6, 4.2, 4.9, and 5.8; and of the evolutionary age
(in Ma) are 0, 85, 111, and 75. We leave the two extinction parameters free: $E(4405-5495)$ (amount of extinction) 
and $R_{5495}$ (type of extinction) and we use the family of extinction laws of \citet{Maizetal14a}. 
See 
\citet{Maiz04c,MaizBarb18}
for an explanation of why one needs to use monochromatic quantities instead of band-integrated ones to characterize extinction. 
We also leave \logd\ free, yielding a total of three 
free parameters to be fitted in each of the four runs. As we are fitting seven photometric points we have four 
degrees of freedom.

{\it Results.} All four runs yield excellent values of the reduced $\chi^2$ around 0.6, indicating that the model
SEDs are consistent with the observed photometry. The values for the extinction parameters are very similar for the
four runs, which was expected because broad-band colors to the right of the Balmer jump are a very weak function of
gravity for B stars. We obtain $E(4405-5495) = 0.413\pm 0.013$ and $R_{5495} = 3.60\pm 0.16$. Note that the value
for the amount of extinction is significantly lower than the one \citet{Liuetal19} obtains but as $R_{5495}$
is also higher than the canonical value of 3.1 there is little change in $A_V$. This deviation from the canonical
extinction law towards higher values is typical of low-extinction environments; only for higher extinctions the
canonical value becomes the more common one \citep{MaizBarb18}, 
which is why it is dangerous to assume a value of $R_{5495}$ instead of actually measuring it. 
The main difference between the four runs are the values of \logd,
which are (in log pc), respectively, 2.975, 3.116, 3.380, and 3.520. Note that for luminosity class 4.5 (third run) we obtain a distance close to 2~kpc (close to the {\it Gaia}~DR2 
value) which corresponds to a value of \logg\ close to 3.6 (cgs), a mass around 5~\MSol, and an age around
110~Ma. Therefore, the photometric-based CHORIZOS analysis leads to a result consistent with the spectroscopic 
analysis in this paper. 

For the second set of runs we use the same photometry as before but we add a Johnson $U$ magnitude by combining the
$U-B$ color from \citet{Mermetal97} with the APASS $B$ magnitude. This allows us to leave \Teff\ as a free
parameter, as the Balmer jump provides an accurate measurement of that quantity for OB stars if the extinction law 
is well characterized 
\citep{Maizetal14a}. 
In this case we do a first run with a luminosity class of 5.0
leaving the two extinction parameters and \logd\ free and a second run where we change the luminosity class to 4.0. 
This leaves in each run four free parameters and eight photometric points i.e. four degrees of freedom.

\begin{figure}
\centerline{\includegraphics*[width=\linewidth]{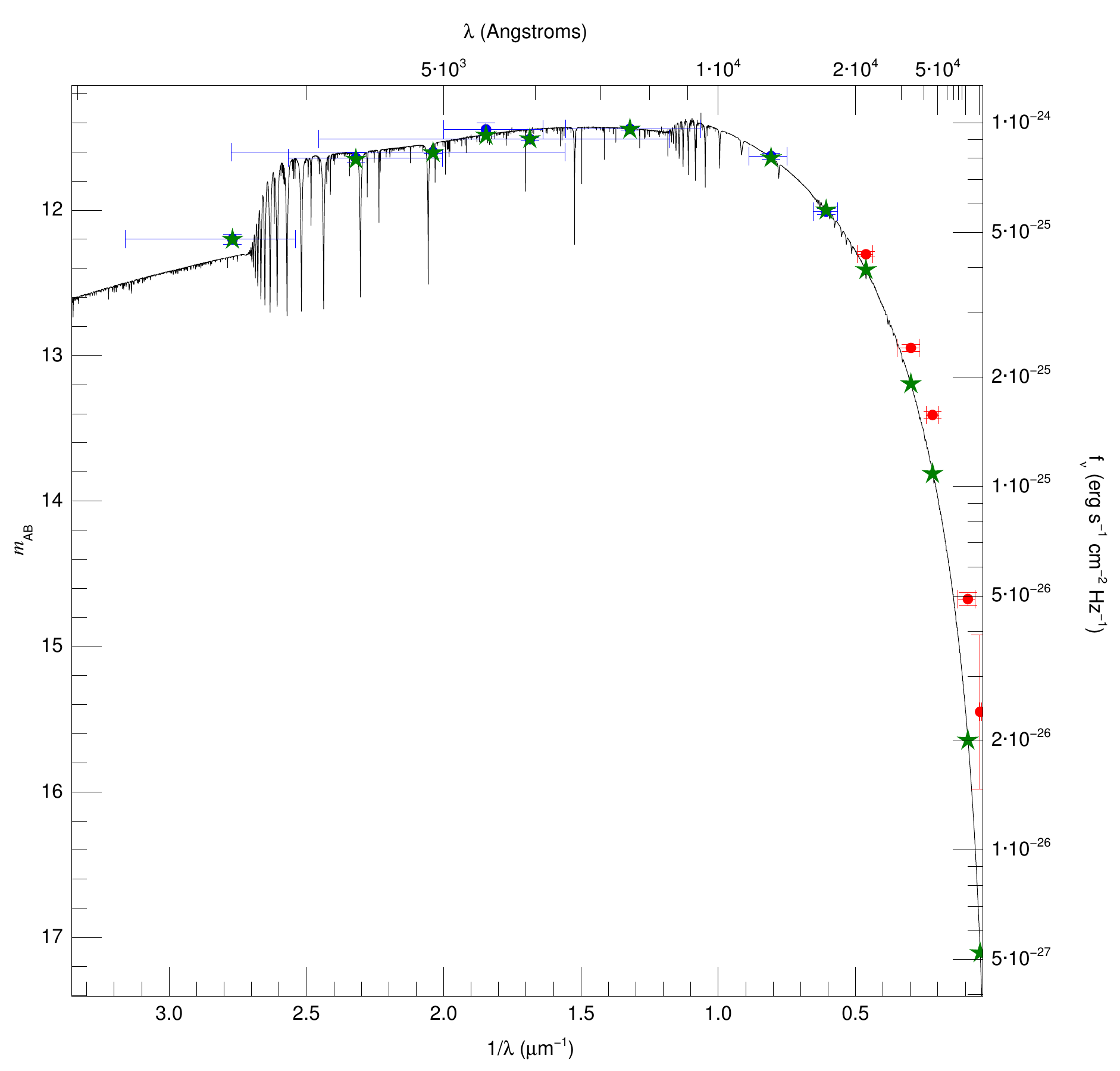}}
\caption{Best-fit SED for the first run of the second CHORIZOS set. Green stars indicate model photometry, blue symbols and error
bars indicate input photometry used for the fit, and red symbols and error bars indicate input photometry not used
for the fit.}
\label{CHORIZOS}
\end{figure}

The first run of the second set has a reduced $\chi^2$ of 0.50. The values of the extinction parameters are within 1.5 sigmas of
the previous ones, $E(4405-5495) = 0.443\pm 0.023$ and $R_{5495} = 3.63\pm 0.14$, and $\logd = 3.219\pm 0.048$~log~pc is
also similar to the previous equivalent result. The important result of this run is that we find 
$\Teff\ = 16\,500\pm 1100$~K, which is relatively close to the value obtained from the spectroscopic analysis and to the value expected for a B6~IV star. 
This implies that the contribution of the disk to the optical continuum is small because if it were not the Balmer
jump would be reduced and we would obtain a significantly hotter temperature. Another piece of indirect evidence in
the same direction are the excellent values of the reduced $\chi^2$ of all runs, as a significant non-stellar
continuum would push the photometric solution out of the parameter space covered by the stellar models. We show in
Fig.~\ref{CHORIZOS} the photometry and best-fit SED. In that Figure it is clear there is a significant infrared
excess as a likely result of the disk contribution. For the best-fit SED the excess is $0.107\pm 0.017$~mag at 
$K_{\rm s}$ and $0.971\pm 0.045$ at $W3$. At $W4$ the input uncertainty is very large but the excess is already
close to 2 mag.

The second run of the second set yields very similar results to the first one except for \Teff\ and \logd. For the first we get 
$15\,700\pm 1100$~K, within one sigma of the previous value and even closer to the spectroscopic one. The small difference indicates, in any case, that the \Teff\ value depends only weakly on the luminosity (or gravity). \logd, on the other hand, is significantly larger
($3.578\pm 0.038$~log~pc), as expected, falling outside the range of the {\it Gaia}~DR2 value.

\section{Gaia DR2 data of LB-1}
\label{app_gaia}

$\,\!$\indent In this Appendix we analyze the {\it Gaia}~DR2 results for LB-1. We start with the parallax $\varpi = 0.4403\pm 0.0856$~mas. The corresponding \citet{Bailetal18} distance is $2140^{+510}_{-310}$~pc. If we use instead the prior described by 
\citet{Maiz01a}
which is more appropriate for early-type stars in the Galactic disk, 
and a parallax zero point of $-0.040\pm0.010$~mas we obtain the alternative similar distance of $2200^{+490}_{-350}$~pc. The proper motion of LB-1 is nearly exactly southwards, as the value in right ascension is $-0.067\pm 0.112$~mas/a and the value in declination is $-1.889\pm 0.088$~mas/a. From the {\it Gaia}~DR2 astrometric point of view, our target is then a typical disk object at a distance of $\sim$2~kpc with a typical proper motion for that population, as those values are close to the average for OB stars at its Galactic longitude \citep{Maizetal18b}. 

Going beyond those basic results, we may ask ourselves: is there a sign of the motion of the B~star around the black hole in the {\it Gaia}~DR2 data? We can use the results from the \citet{Liuetal19} spectroscopic SB1 orbit to calculate the motion in the plane of the sky. For an edge-on ($\sin i=1$) orbit, the B star moves back and forth along a line with a semi-amplitude of $a_0 = 0.19\,(2\,{\rm kpc}/d)$~mas, that is, 38\% of the amplitude caused by Earth's motion as measured in {\it Gaia}~DR2. If the orbit is not edge-on, the trajectory is an ellipse with semi-major axes of $a_0/\sin i$ and $a_0/\tan i$, respectively. Note that for a star with $G=11.4$ the precision for an along-scan position measurement is $\approx$0.3~mas \citep{Lindetal18a}. From those values it is clear that already for the edge-on case one should see an effect of the motion of the B star around the black hole. For an orbit with a smaller inclination the effect should be even larger and for low enough inclinations it should dominate over the effect of the Earth's motion. However, as the orbital motion is in an arbitrary direction with respect to that of the Earth's orbit and the period is much shorter than one year, the effect in the DR2 astrometric solution should appear as noise, not as signal. In order to actually measure the orbit of the B star around the black hole using {\it Gaia} data one needs access to the individual epochs, something that is not available in the current release but that will become available in the future (see e.g. \citealt{Andretal19}). Therefore, for the time being we have to work with the available limited information.

Despite our calculations in the previous paragraph, {\it there is no sign of such an orbital motion in the Gaia DR2 data.} We can see this in two ways. First, the RUWE parameter \citep{Lindetal18b}, which is the recommended goodness-of-fit for {\it Gaia}~DR2 astrometry, is 0.91, close to the ideal value of 1 and much lower than the suggested critical value of 1.4. Second, {\it Gaia}~DR2 attempted to use 103 along-the-scan observations and only discarded three, indicating that there were very few outliers to the fit\footnote{A possible third way to detect this would be through the excess astrometric noise but that should not be used for a bright star like this one due to the problems associated with the ``DOF bug'', see appendix A in \citet{Lindetal18a}.}. Therefore, we face a conundrum: if the only  significant contribution to the $G$ band photometry, and hence astrometry, in the system comes from the B star and the \citet{Liuetal19} spectroscopic orbit is correct, what is going on with the {\it Gaia} astrometry? We believe that the issue has to do with the special circumstances of this system. First, LB-1 is located very close to the northernmost point of the ecliptic, which results in the reflex motion of the star due to the Earth's motion being in an east-west line instead of in a true ellipsoid. Second, the proper motion of the star is in a nearly perpendicular direction (close to the north-south direction) to that. Therefore, ignoring the effect of the BH the trajectory of the star in the sky is close to a sinusoid, as opposed to the complex trajectories that usually result of the sum of a linear motion with an arbitrary velocity (proper motion) and an ellipse whose shape is determined by the position in the sky and its amplitude by its parallax. In other words, it is a case where the fit parameters are degenerate and subject to peculiar results when perturbations are included. In this particular case, we propose that {\it if the B-star orbit has indeed a very high inclination with its axis pointing in a near east-west direction, Gaia would detect little excess noise in the parallax measurements} and would only see the star slowing down and speeding up in the proper motion direction, something that may not come up as excess noise in the output. The most critical aspect is the high inclination, as a lower value would make the orbit span a larger angular size in the sky that would be harder to miss by {\it Gaia}. In this respect, we point out that even though in {\it Gaia}~DR2 there are 12 epochs observed, three of them are actually close in time to others and there are only nine real visibility periods employed in the solution, so it is not far-fetched to think that the orbital motion would have been missed if the system is in the proposed configuration. The answer should come in future releases when more epochs are included in the solution and information from individual epochs becomes available. In the meantime, it would be useful to make polarization observations in H$\alpha$ to see if the proposed circumbinary disk is indeed oriented this way.

If the hypothesis in the paragraph above (or a slightly modified version of it) turns out to be true, there are two interesting consequences for LB-1. The first one is that a high inclination in the spectroscopic orbit results in a significantly lower mass for the black hole. If the B star has a mass of 5~\MSol, for $\sin i = 1$ we are left with a black hole of only 4.9~\MSol\ (for $\sin i = 0.7$ its mass becomes 8.6~\MSol\ and for $\sin i = 0.4$, 26.1~\MSol). If the B star has instead a mass of 3~\MSol, the minimum mass of the BH is just 3.8~\MSol. The second one is that, as we have measured a very low rotational $v\sin i$ for the B star, either $v$ is really very low (the star is a true slow rotator) or its rotation axis points in a near-perpendicular direction with respect to its orbital axis.

\end{appendix}

\end{document}